\documentclass[aps,prd,preprint,groupedaddress,preprintnumbers,floatfix]{revtex4}

\usepackage{amssymb}
\usepackage{amsmath}
\usepackage{verbatim}
\usepackage{mathrsfs}
\usepackage{amsfonts}
\usepackage{latexsym}
\usepackage{epsfig}
\usepackage{color}
\usepackage{cancel}
\usepackage{graphicx,subfigure}
\usepackage{units}

\begin{document}


\title{Constraining New Physics with $D$ meson decays}

\author{J. Barranco}
\email{jbarranc@fisica.ugto.mx}

\author{D. Delepine}
\email{delepine@fisica.ugto.mx}
\author{V. Gonzalez Macias}
\email{vanniagm@fisica.ugto.mx}
\author{L. Lopez-Lozano}
\email{lao-tse@fisica.ugto.mx}

\affiliation{Departamento de F\'isica, Divisi\'on de Ciencias e Ingenier\'ia, Campus Le\'on,
Universidad de Guanajuato, Le\'on 37150, M\'exico}

\begin{abstract}
Latest Lattice results on $D$ form factors evaluation from first principles show that the standard model (SM) 
branching ratios prediction for the leptonic $D_s \to \ell \nu_\ell$ decays and the semileptonic SM branching 
ratios of the $D^0$ and $D^+$ meson decays are in good agreement with the world average experimental measurements. 
It is possible to disprove New Physics hypothesis or find bounds over several models beyond the SM. 
Using the observed leptonic and semileptonic branching ratios for the D meson decays, we performed a combined 
analysis to constrain non standard interactions which mediate the $c\bar{s}\to l\bar{\nu}$ transition.
This is done either by a model independent way through the corresponding Wilson coefficients
or in a model dependent way by finding the respective bounds over the relevant parameters for some models beyond 
the standard model. In particular, we obtain bounds for the Two Higgs Doublet Model Type-II and Type III, 
the Left-Right model, the Minimal Supersymmetric Standard Model with explicit R-Parity violation and
Leptoquarks. 
Finally, we estimate the transverse polarization of the lepton in the $D^0$ decay and we found it can be as high as $P_T=0.23$.
\end{abstract}


\maketitle

\section{Introduction}
In spite of the Standard Model (SM) success, now favored by the probable recent discovery of the Higgs 
boson \cite{:2012gk,:2012gu},
the search of a more fundamental theory at an energy scale much bigger than the electroweak scale is still open.
Interestingly, low energy scale experiments may shed some light in the search for such fundamental theory due to their
possibility of getting high statistics and hence indirect observables of New Physics (NP).
We will use D meson decays as an illustration.
Contrary to $B$ meson physics, charmed hadronic states are in the unique mass
range of $O(2\,\mathrm{GeV})$, which allows for strong non perturbative hadronic physics
\cite{Ryd:84.65}. Moreover, the calculations for the relevant form factors, which parameterize all QCD effects within the hadronic state, have been
improved significantly reaching a remarkable precision \cite{Davies:2010ip,Aubin:2004ej,Koponen:2013tua}. The SM predictions for the D meson leptonic and semileptonic decays relies on the lattice QCD estimates of the form factors, and appear to be in agreement with the world average experimental measurements \cite{Koponen:2013tua}, allowing us to disprove New Physics hypothesis or find restrictive bounds over several models beyond the SM.

\begin{figure}
\includegraphics[width=0.9\textwidth]{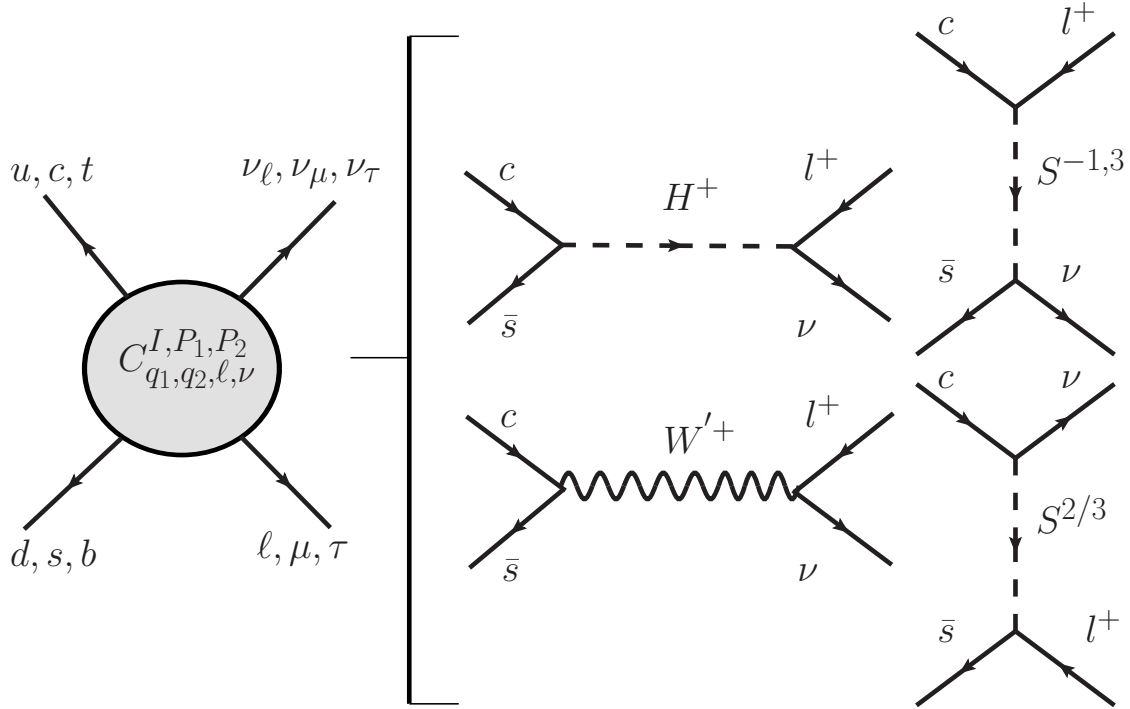}
\caption{Generic charged current non standard interaction between two quarks and the leptonic sector.
Some Feynman diagrams for models beyond SM involving the $c\bar{s}\to l\bar{\nu}$ transition involved in
D meson decays are shown.}\label{NSIgeneric}
\end{figure}

At low energies, most of the extensions to the Standard Model reduce to an effective four Fermi interaction,
usually called Non Standard Interaction NSI, that can be parameterized by a generic coefficient (Fig. \ref{NSIgeneric}).
For the $\Delta C=\Delta S$ leptonic and semileptonic D meson decays, the new particle state should couple to the leptons
and the second generation of quarks, leaving such effective interaction. Any kind of intermediate state,
such as scalars, vectors or even tensors, are allowed. Examples are the Two Higgs Doublet Model Type-II (THDM-II) and Type III (THDM-III)
\cite{Donoghue:1978cj}, the Left-Right model (LR)) \cite{Pati:1973rp,Mohapatra:1974hk,Mohapatra:1974gc,Senjanovic:1975rk,Senjanovic:1978ev}, the Minimal Supersymmetric Standard Model with
explicit R-Parity violation (MSSM-$\cancel{R}$)  \cite{Dreiner:1997uz,Barbier:2004ez}, and the Leptoquark model \cite{Pati:1974yy,Buchmuller:1986zs}, also illustrated in
Fig. \ref{NSIgeneric}.

Non Standard interactions from a model independent approach had been considered and constrained with $D_s$ leptonic decays \cite{Dobrescu:2008er,Carpentier:2010ue},
and independently, using semileptonic decays \cite{Kronfeld:2008gu,Carpentier:2010ue}.  In this work we make a model independent analysis and a model dependent analysis in order to constrain NSIs combining the leptonic and semileptonic decays of the D meson. We use the latest Lattice results on the form factors\cite{Koponen:2013tua} which have reached a significant precision. We show the usefulness of the model independent constraints as well as specific cases when a model dependent analysis is needed. The $q^{2}$ distributions for the $D^{+}\to \bar{K}^{0} e^{+}\nu_{e}$ and  $D^0 \rightarrow K^- e^{+} \nu$ decays, which are expected to be sensitive to new physics, are also considered.
 Using the respective bounds for the Wilson coefficients, we compute as well the transverse polarization of the charged lepton in the semileptonic decay of the D meson. This $T$ violating observable has not been measured but may provide significant constraints over the complex character of the new physics parameters, as in the case of the B meson semileptonic decay \cite{Tanaka:2012nw} and other meson decays \cite{Abe:2006de}. The paper is organized as follows: In Section \ref{s2} we describe the general effective Lagrangian for the semileptonic transition $c\to s$ when non standard interactions are included and show the theoretical branching ratios and the transverse polarization of the D meson semileptonic decay. In Section \ref{s3} we show the experimental constraints over the Wilson coefficients, and the theoretical predictions for the transverse polarization of the D meson semileptonic decay. In Section \ref{s4} we constrain the relevant parameters of the THDM-II and THDM-III, LR, and the MSSM-$\cancel{R}$, and the leptoquark model. In Section \ref{s5} we give our conclusions and comments over the relevance of these bounds.

\section{Non standard interactions and relevant observables}\label{s2}
\subsection{Effective Lagrangian below the electroweak scale}
The search of new physics effects in the leptonic and semileptonic processes of mesons has two sources of
uncertainty that can not be separated; the non perturbative long-distance forces that bind quarks forming
hadronic states and the determination of the free parameters of the SM, \emph{i.e.} fermions masses and
CKM matrix elements. The non-perturbative QCD effects are parameterized introducing form factors.
On the other hand effective couplings that correspond to short distance interactions could receive non-standard contributions.
Hence flavour-changing meson transitions in the SM have at least two scales involved, the electroweak scale that is responsible of the flavour changing and the scale of strong interactions \cite{Antonelli:2009ws}. When NSI are considered, we assume that the new physics energy scale is higher than the electroweak scale, thus the operator product expansion formalism (OPE) \cite{Wilson:1969zs} is suitable since it allows the separation between long-distance (low energy) and short-distance (high energy) interactions. In the OPE the degrees of freedom corresponding to higher energies scales are integrated out \cite{Gaillard:1974nj}, resulting an effective Lagrangian where all high energy physics effects are parameterized by Wilson's coefficients, namely the effective couplings multiplying the operators of the Lagrangian.
In this spirit, the non-standard effective Lagrangian for a semileptonic transition as the one
illustrated in Fig. \ref{NSIgeneric} is:
\begin{equation}\label{effec_lagrangian}
-\frac{\mathcal{L}_{NP}}{G_F}=\sum_{\substack{c,s,\ell,\nu\\I=S,V,T\\P_{1,2}=L,R}}C^{I,P_1P_2}_{q_1q_2\ell\nu}(\bar{q_1}\Gamma^IP_1 q_2)\cdot(\bar{\nu}_L\Gamma_IP_2\ell)\,,
\end{equation}
where the indexes $q_1$ and $q_2$ represent down-type and up-type quarks respectively, $\ell$ is the charged lepton
flavor and $\nu$ its corresponding neutrino. $P_{1,2}$ represent the chiral projectors $L=(1-\gamma^5)/2$ and
$R=(1+\gamma^5)/2$. Here, the current operators $\Gamma$'s are determined by the Dirac field bilinears,
namely: $\Gamma_S=1$, $\Gamma_V=\gamma_\mu$ and $\Gamma_T=(i/2)[\gamma^\mu,\gamma^\nu]$.
The dimensionless coefficients $C^{I,P_1P_2}_{q_1q_2\ell\nu}$ have a clean interpretation: they are a measurement of
how big can the NSI be as compared to the SM current, since they are weighted by the Fermi constant $G_F$.
This parametrization technique enables us to test NSI when the experiments reach certain precision, and in particular
to look for NP effects at low energies.

\subsection{NSIs in the $D$ meson leptonic and semileptonic decays}
The decay rate of $D_s\to\ell\nu_\ell$ including the SM Lagrangian plus the NSI Lagrangian of  eq. (\ref{effec_lagrangian}),
is thus given by
\begin{eqnarray}
&&\Gamma_{D_s\to\ell\nu}=\frac{|G_{F}f_{D_s}\left(M^{2}_{D_s}-m^{2}_{l}\right)|^2}{8\pi M_{D_s}^3}\left|V_{cs}m_l +
\right.\nonumber\\
&&\left.
\frac{m_{l}(C^{V,LL}_{sc\ell\nu}-C^{V,RL}_{sc\ell\nu})}{2\sqrt{2}}+\frac{M^{2}_{D_s}(C^{S,RR}_{sc\ell\nu}-C^{S,LR}_{sc\ell\nu})}{2\sqrt{2}(m_c+m_s)}
\right|^2\,.\label{gamma_leptonic}
\end{eqnarray}

On the other hand, in the rest frame (RF) of the decaying meson, the partial decay rate for the $D^{0}\to K^{\pm} l^{\mp} \nu$ decay channel with non standard interactions is given by

\textbf{\begin{eqnarray}
&&\frac{d\Gamma_{D\to K\ell\nu_\ell}}{dE_K}=
\frac{G_F^2 m_D \sqrt{E_K^2-m_K^2}}{(2\pi)^3}
\left\{ (E_K^2-m_K^2) \frac{2q^2 + m_\ell^2}{3q^2}
\left|G_Vf_+(q^2)\right|^2 \right.
\nonumber
\\ &+& \left.\left(-|G_Tf_2(q^2)|^2\frac{q^2+2m_l^2}{3}+
m_lG_Vf_+(q^2)G_T^*f_2(q^2)\right)\left(\frac{E_k^2-m_K^2}{m_D^2}\right)
\right. \nonumber\\ &+&
\left.\frac{\left|(m_D^2-m_K^2)qf_0(q^2)\right|^2}{4m_D^2}
\left|\frac{m_\ell}{q^2} G_V+\frac{G_S}{m_c-m_s}\right|^2
\right\} \left(1-\frac{m_\ell^2}{q^2} \right)^2\,,\label{gamma_semileptonic} 
\end{eqnarray}}

where in the later expression we have defined
$G_V=V^{*}_{cs}+(C^{V,LL}_{sc\ell\nu}+C^{V,RL}_{sc\ell\nu})/2\sqrt{2}$, 
$G_S=(C^{S,RR}_{sc\ell\nu}+C^{S,LR}_{sc\ell\nu})/2\sqrt{2}$ and $G_T=(C^{T,RR}_{sc\ell\nu}+C^{T,LR}_{sc\ell\nu})/2\sqrt{2}$.
Other constants involved
in Eqs. (\ref{gamma_leptonic}, \ref{gamma_semileptonic}) are:
$G_F$ the Fermi constant, $V^{*}_{cs}$ the CKM matrix element, $m_\ell,m_c,m_s,m_K,m_{D_s},m_D$ the masses of the
leptons, charm and strange quarks, the Kaon and D meson respectively as reported by PDG \cite{Beringer:1900zz}.
The transferred energy is $q^2=m_D^2+m_K^2-2m_DE_K$ and $E_K$ is the final energy of the Kaon meson. Its allowed energy is
$m_K<E_K<(m_D^2+m_K^2-m_\ell^2)/2m_D$.
The decay constant $f_{D_s}$ in the leptonic decay rate is defined by 
$\langle0|\bar s\gamma_\mu\gamma_5c|D_s(p)\rangle=if_{D_s}p_\mu$.
In the semileptonic decays, the scalar and vectorial form factors $f_0(q^2)$ and $f_+(q^2)$ are defined 
via $\langle K|\bar s\gamma^\mu c|D\rangle=f_+(q^2)(p_D+p_K-\Delta)^\mu+f_0(q^2)\Delta^\mu$, 
with $\Delta^\mu=(m_D^2-m_K^2)q^\mu/q^2$, and $\langle K|\bar s c|D\rangle=(m_D^2-m_K^2)/(m_c-m_s)f_0(q^2)$. 
\subsection{Transverse polarization including NSIs}
The transverse polarization of the charged lepton in the decay $D\to K l \nu$ is a sensitive 
T-violating or CP violating observable when CPT is conserved. 
This observable was first computed in the semileptonic decay $K^+ \to \pi^0 \mu^+ \nu$ as a useful tool for studying 
non standard CP violation \cite{Leurer:1988au,Garisto:1991fc} .
In fact, within the SM, this transverse polarization is identically zero. Therefore, a non zero value is a 
clear signal of new physics. 
Given the similarities with the $K^+$ decay, we can compute the transverse polarization for the
semileptonic decay of the D meson $D(p)^{0}\to K^{\mp}(k)\nu(p_1)l(p_2)^{\pm}$.
The transverse polarization is given by
\begin{equation}
P_T^S=\frac{|A_T^S|^2-|A_T^{-S}|^2}{|A_T^S|^2+|A_T^{-S}|^2}
\end{equation}
where $S$ represents the spin of the lepton. In general, one measures the spin perpendicular to the decay plane defined 
by the final particles \cite{Abe:2006de}. Thus in order to have a non zero effect the transverse polarization should 
be proportional to $\epsilon^{\alpha\beta\gamma\delta}p_\alpha p_{1\beta}p_{2\gamma}S_\delta$, where $p_1$ and $p_2$ are the 
4-vectors of neutrino and charged lepton respectively. Given that $S_\mu=(0,\mathbf{s})^T$, with $\mathbf{s}$
perpendicular to the decay plane, the polarized amplitude can be written as
\begin{align}
|A_T^S|^2&=\frac{1}{2}|A_{D\to K\ell\nu_\ell}|^2\nonumber\\
+& 8G_F^2\epsilon^{\alpha\beta\gamma\delta}K_\alpha S_\beta p_{1\gamma}p_{2\delta}
\left[f_+(q^2)f_0(q^2) \frac{M_D^2-M_K^2}{m_c-m_s} \textrm{Im}(G_VG_S^*)\right. \nonumber \\
+& \left. f_2(q^2)\left((f_0(q^2)-f_+(q^2))(M_D^2-M_K^2)(1-2\frac{p_2\cdot q}{q^{2}})\right.\right.\nonumber \\
+&\left.\left. f_+(q^2)\frac{q\cdot Q-2p_2\cdot Q}{M _D}\right)\textrm{Im}(G_V G_T^*) \right.\nonumber \\
+&\left. f_2(q^2) f_0(q^2)\frac{m_\ell}{M_D}\frac{M_D^2-M_K^2}{m_c-m_s}\textrm{Im}(G_TG_S^{*})\right],
\end{align}
here $Q=p+k$ and $q=p-k$.
With this, we can construct the transverse polarization averaged over the charged lepton energy. 
To calculate the averaged transverse polarization
we have integrated over the  charged lepton energy.
Thus this observable can be written in the decay frame of the $D$ meson as

\begin{eqnarray}
\langle P_T^S\rangle &=&\frac{G_F^2 M_D^4}{4 \pi^3}
\left(\frac{d\Gamma}{dE_K}\right)^{-1}
\left\{f_0(q^2)f_+(q^2)\frac{M_D^2-M_K^2}{M_D(m_c-m_s)} g_0(q^2)\textrm{Im}(G_V G_S^{*}) \right. \nonumber \\
&+&\left. f_2(q^2)f_0(q^2)\frac{m_\ell}{M_D^2} \frac{M_D^2-M_K^2}{m_c-m_s} g_0(q^{2})\textrm{Im}(G_T G_S^{*})\right. \nonumber\\ 
&+&  \left. f_2(q^2)\left[f_0(q^2)\frac{m_\ell^2}{M_D^2}\frac{(M_D^2-M_K^2)}{q^2}g_0(q^2)
+f_+(q^2)\left(g_1(q^2) \right.\right.\right.\nonumber \\
&-&\left. \left.\left. \frac{m_\ell^2+q^2}{M_D^2}\frac{(M_D^2-M_K^2-q^2)}{q^2}g_0(q^2)\right)\right]\textrm{Im}(G_T G_V^{*})\right\}
\label{pol}
\end{eqnarray}

where we have defined the dimensionless kinematical functions

\begin{equation}
g_0(q^2)\equiv\frac{1}{M_D^3}\int_{E_\ell^{\textrm{min}}}^{E_\ell^{\textrm{max}}} dE_\ell |\mathbf{p_1}\times\mathbf{p_2}|\,,\qquad
g_1(q^2)\equiv\frac{1}{M_D^4}\int_{E_\ell^{\textrm{min}}}^{E_\ell^{\textrm{max}}} dE_\ell E_\ell \cdot |\mathbf{p_1}\times\mathbf{p_2}|\,,
\end{equation}
with
\begin{equation}
E_\ell^{\textrm{max (min)}}=\frac{1}{2}\left(m_D-E_K\right)\left(\frac{m_\ell^2}{q^2}+1\right)
\pm \frac{1}{2}\sqrt{E_K^2-m_K^2}\left(1-\frac{m_\ell^2}{q^2}\right)\,.
\end{equation}
We can see that the leading contributions in New Physics are the scalar and tensor interactions, i.e. at first order in $C$'s.

\section{Model independent analysis and experimental constraints}\label{s3}
\subsection{$D$ meson decays measurements {\it vs} theoretical branching ratios}
\begin{table}[t]
\begin{center}
\caption{Theoretical and experimental branching ratios}\label{table1}
\begin{tabular}{c|c|c|c}
\hline\hline
$i$&Decay& Theoretical BR $\mathcal{B}_i^{th}$ & Experimental BR $\mathcal{B}_i^{exp}$ \\
\hline\hline
1&$D^0\to K^- e^+ \nu_e$            & $(3.28\pm0.11)\%$.          & $(3.55\pm0.04)\%$\\
2&$D^0\to K^- \mu^+ \nu_\mu$        & $(3.22\pm0.11)\%$           & $(3.30\pm0.13)\%$\\
3&$D^+\to \bar K^0 e^+ \nu_e$       & $(8.40\pm0.32)\%$.          & $(8.83\pm0.22)\%$\\
4&$D^+\to \bar K^0 \mu^+ \nu_\mu$   & $(8.24\pm0.31)\%$           & $(9.2\pm0.6)\%$\\
5&$D_s^+\to \tau^+ \nu_\tau$         & $(5.10\pm0.22)\%$          & $(5.43\pm 0.31)\%$\\
6&$D_s^+\to \mu^+ \nu_\mu$          & $(5.20\pm0.20)\times10^{-3}$ & $(5.90\pm 0.33)\times10^{-3}$\\
\hline\hline
\end{tabular}
\end{center}
\end{table}

There are a number of measurable observables related to the D meson that might be  modified by NSI. 
$D_s$ leptonic decays have been measured by a number of experiments, namely
CLEO \cite{Besson:2007aa} and Belle \cite{Widhalm:2007ws} among other experiments.
Semileptonic decays, on the other hand, have been observed with an integrated luminosity of 
818pb$^{-1}$ \cite{Link:2004dh,Aubert:2007wg,Dobbs:2007aa}. In particular,
the $q^{2}$ distribution for the semileptonic decays $D^+\to \bar K^0 e^+ \nu_e$ , $D^0\to K^- e^+ \nu_e$ 
has been measured by CLEO \cite{Besson:2009uv},\cite{Ge:2008aa}. 
From those measurements it is possible to extract the lifetimes for the mesons. They result to be
$\tau_{D^0}=(410.1\pm1.5)\times 10^{-15}~$s,  $\tau_{D^+}=(1040 \pm 7)\times 10^{-15}~$s, and
$\tau_{D_s}=(500 \pm 7)\times 10^{-15}~$s.
In summary, total branching ratios for semileptonic decays of the $D^0$ and $D^+$  and
the world measured total branching ratios for the leptonic decays of the $D_s$ are shown in
Table \ref{table1}.\\
The theoretical decay rates, on the other hand, 
$\Gamma_{D_s\to\ell\nu}^{th}$, $\Gamma_{D^0\to K^+\ell^-\nu_\ell}^{th}$ and  $\Gamma_{D^+\to \bar K^0\ell^+\nu_\ell}^{th}$
given by eqs. (\ref{gamma_leptonic},\ref{gamma_semileptonic}) are computed by fixing all the Wilson's coefficients to zero. 
We ignore all radiative corrections since they are expected to be below the 1\% \cite{Burdman:1994ip}. 

Other relevant physical inputs needed for the SM computation of the theoretical BRs are:
\begin{enumerate}
\item {\it The CKM element $V_{cs}$.} As we are looking for New Physics, we have to be very careful on the value of the CKM element we will use in our numerical analysis. In order to avoid that leptonic and semi leptonic of $D$ mesons have been used to fix  the $V_{cs}$ value, we use the central value of the CKM element which comes from $W \to cs$ decay, neutrino-nucleon scattering and unitary constraints coming from $b-s$ transitions relating $|V_{cd}|$ and $|V_{cs}|$ through unitarity. This last constraint gives the strongest constraint. So our central value for  $V_{cs}$ is $0.97344 \pm 0.00016$ 
\cite{Charles:2004jd}. 
Using this unitary constraint means that automatically our results will not apply to any model with more than three fermion families. 

\item {\it Hadronic form factors.}  These are non-perturbative parameters calculated in specific theoretical models. In particular Lattice QCD is a well-established method able to compute the hadronic form factors from first principles, that has reached an excellent precision \cite{Koponen:2013tua}. Therefore, for our analysis, we fix the hadronic form factors and leptonic decay constant to the value estimated with lattice QCD simulations. 
The leptonic decay constant $f_{D_s}$ has been computed with a precision of the order of 2\% by the 
HPQCD collaboration \cite{Davies:2010ip}. In order to compute the leptonic branching ratio we have used
the reported value of $f_{D_s}=248\pm2.5~$MeV \cite{Davies:2010ip}.
On the other hand, less is known about $f_0(q^2),f_+(q^2)$. Dramatic progress has been made over the last decade on 
lattice calculations of for those form factors \cite{Davies:2010ip,Aubin:2004ej,Koponen:2013tua}. We use the latest
results by the HPQCD collaboration \cite{Koponen:2013tua} as input for the calculation of the theoretical decay rate.

\end{enumerate}

\begin{figure}
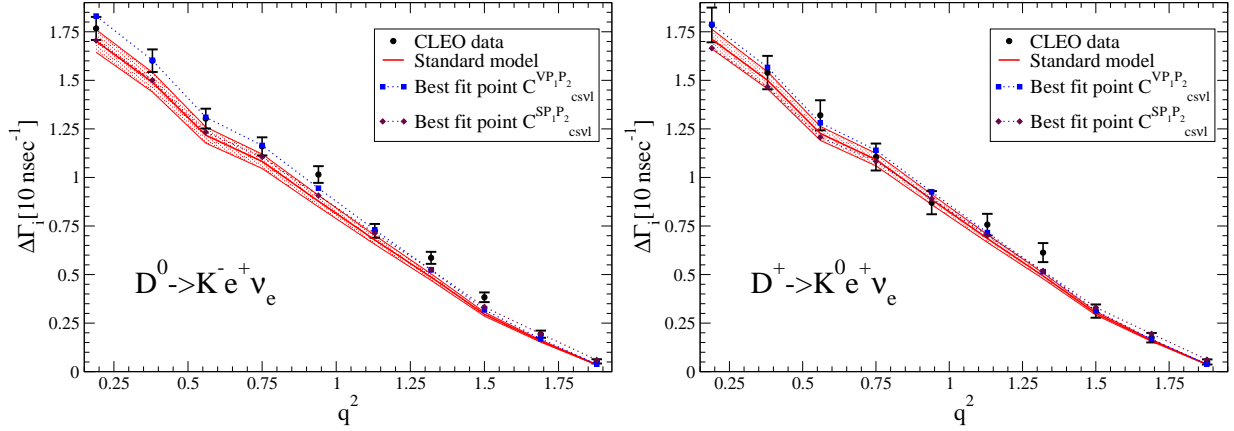

\includegraphics[width=0.49\textwidth]{gamma_theo_exp.eps}
\includegraphics[width=0.49\textwidth]{gammaDM_theo_exp.eps}
\caption{Partial decays measured by CLEO \cite{Ge:2008aa} and the theoretical partial decay computed with the Standard model using the
latest form factors from \cite{Koponen:2013tua}. Grey region represents one sigma theoretical error. Good agreement is observed}\label{partialdecay}
\end{figure}

The results for the theoretical BRs are listed in table \ref{table1}. with their corresponding uncertainties. 
The total theoretical uncertainties are calculated straightforward: propagating each uncertainty for every physical 
constant as reported in PDG\cite{Beringer:1900zz}, and the theoretical uncertainties coming from the lattice QCD 
calculations of the form factors. The main contribution in the theoretical error comes from the leptonic decay constant $f_{D_s}$ and the semileptonic form factors $f_{+}(q^2)$ and $f_{0}(q^2)$. The reported error in $f_{D_s}$ 
induces a $\sim 4\%$ error in the theoretical leptonic branching
ratio. Similarly, the reported error in the lattice determination of $f_0(q^2)$ and $f_+(q^2)$ leads to a
$\sim 4\%$ error in the theoretical semileptonic branching ratio. Exact values are listed in table \ref{table1}.
As already mentioned, world average measurements of the total BRs as reported by PDG \cite{Beringer:1900zz} are shown in 
Table \ref{table1} for comparison. In the same way, the theoretical partial decays for the $D^0\to K^+e^-\nu_e$ 
and $D^+\to \bar K^0 e^+\nu_e $ and the CLEO data points are shown in Figure \ref{partialdecay}. 
Note the good agreement between experiment and theory.\\

The BRs reported in table \ref{table1} are a pure theoretical
prediction of the SM in the following sense: $\mathcal{B}^{th}$ is computed using
the SM Lagrangian only, since we have set all Wilson coefficients to zero,
and the form factors are computed from first principles using Lattice results\cite{Koponen:2013tua}.

\subsection{Constraining real NSI }
\begin{table*}
\begin{center}
\begin{tabular}{|c|c|c|c|c|}
\hline
&4 pars. 95\% C.L.&$\chi^2_{min}/\mbox{d.o.f}$ & 1 par. 95\% C.L.& $\chi^2_{min}/\mbox{d.o.f}$\\
\hline
$C^{V,LL}_{sc\ell\nu}$& $[-.094,0.42]$ &$0.62$ & $[0.072,0.14]$& $0.89$    \\
$C^{V,RL}_{sc\ell\nu}$& $[-0.34,0.17]$ &$0.62$ & $[0.057,0.13]$& $1.29 $\\
$C^{S,RR}_{sc\ell\nu}$& $[-0.33,0.21]$ &$0.62$ & $[-0.22,-0.21]$ $\cup$ $[0.00,0.13] $ &  $2.19$ \\ 
$C^{S,LR}_{sc\ell\nu}$& $[-0.23,0.33]$ & $0.62$ & $[-0.012,0.00]$ $\cup$ $[0.20,0.22] $ &$2.17$ \\
\hline
\end{tabular}
\end{center}
\caption{Model independent constraints at 95\% C.L. for universal non standard interactions using
leptonic and semileptonic D meson decays. We have fixed the leptonic decay constant and semileptonic
form factors to those estimated by lattice QCD. In the first column, four parameters are allowed to vary at a time and in the third column, only one parameter is varied.}\label{table2}
\end{table*}
\begin{table*}
\begin{center}
\begin{tabular}{c}
Flavor dependent scalar non standard interactions\\
\end{tabular}
\end{center}
\begin{center}
\begin{tabular}{|c|c|c|c|c|}
\hline
& 95\% C.L. $\Re^{+}$ &$\frac{\chi^2_{min}}{\mbox{d.o.f}}$ &  95\% C.L. $\Re$ &$\frac{\chi^2_{min}}{\mbox{d.o.f}}$ \\
\hline
$C^{S,RR}_{sc e \nu_e}+C^{S,LR}_{sc e\nu_e}$& $[0.32,0.47]$& $1.05$& $[-0.47,-0.33]\cup[0.32,0.47]$&$1.05$\\
\hline
$C^{S,RR}_{sc \mu \nu_\mu}$& $[0.0,0.27]$ & $1.17$
& $[-0.77,0.25]$& $1.30$\\
$C^{S,LR}_{sc \mu \nu_\mu}$& $[0.0,0.38]$ & $1.17$ &
$[-0.63,0.38]$& $1.30$\\
\hline
$C^{S,RR}_{sc \tau \nu_\tau}-C^{S,LR}_{sc\tau\nu_\tau}$& &  & $[-0.075,0.175]$ & $1.0$\\
\hline
\end{tabular}
\end{center}
\begin{center}
\begin{tabular}{c}
Flavor dependent vector non standard interactions\\
\end{tabular}
\begin{tabular}{|c|c|c|c|c|}
\hline
&  95\% C.L. $\Re^{+}$ &$\chi^2_{min}/\mbox{d.o.f}$ &  95\% C.L. $\Re$ & $\chi^2_{min}/\mbox{d.o.f}$\\
\hline
$C^{V,LL}_{sc e \nu_e}+C^{V,RL}_{sc e\nu_e}$& $[0.07,0.14]$ & $0.99$ & $[0.07,0.14]$ & $0.99$ \\
\hline
$C^{V,LL}_{sc \mu \nu_\mu}$ & $[0.0,0.22]$ &$1.67$ &$[-0.025,0.255]$& $0.93$ \\
$C^{V,RL}_{sc \mu \nu_\mu}$& $[0.0,0.1]$ &$1.67$ &$[-0.19,0.095]$& $0.93$  \\
\hline
$C^{V,LL}_{sc \tau \nu_\tau}-C^{V,RL}_{sc\tau\nu_\tau}$& & & $[-0.12,0.28]$&$1.0$ \\
\hline
\end{tabular}
\end{center}
\caption{Model independent constraints at 95\% C.L. for scalar and vector flavor dependent non standard interactions from the leptonic and semileptonic D meson decays. We have fixed the leptonic decay constant and semileptonic
form factors to those estimated by lattice QCD. In the second column, the Wilson coefficients are restricted to be positive. In the fourth column, the Wilson coefficients are only restricted to be real numbers.}\label{table3}
\end{table*}

Let us assume that the new physics effects, are parameterized, as described in section \ref{s2}, 
by the Wilson coefficients. In this first part of our analysis we suppose the non standard physical phases are aligned with those of the SM in such a way that in general we can consider the Wilson coefficients real. We compute the range of the Wilson coefficients to exactly match the theory 
and the experiment. In order to do so, we perform a simple $\chi^2$ analysis, 
with $\chi^2=\sum_i (\mathcal{B}^{th}_i-\mathcal{B}^{exp}_i)^2/
\mathcal{\delta B}_i^2$. Here, $\mathcal{\delta B}_i$ is calculated adding in quadratures the experimental
and theoretical uncertainties shown in Table \ref{table1}.\\

We shall consider first a combined analysis of the leptonic and semileptonic BRs and the experimental data
from CLEO assuming only 
scalar (S) and vector (V) NSI. An analysis including all the New Physics operators at a time, scalar, vector and tensor, shows that the tensor contribution is negligible as compared to the former operators.  However we can constrain the tensor interactions assuming that only the tensor operator is dominant, as we show in the next subsection.
Hence, the relevant parameters with the above considerations are:
$C^{V,LL}_{sc\ell\nu},C^{V,RL}_{sc\ell\nu},C^{S,RR}_{sc\ell\nu}$ and $C^{S,LR}_{sc\ell\nu}$. 
Although this is a restrictive hypothesis, this analysis is useful for models where no CP violating phases
or models in which the phyiscal phases are aligned with the CKM phase, e.g THDM-II or some 
specific MSSM-$\cancel{R}$ as we will show later.
The results for the relevant Wilson coefficients, assuming these are flavor universal or flavor dependent, are shown in tables \ref{table2} and \ref{table3} respectively.

\begin{itemize}
\item{\it Flavor independent NSI}
Table \ref{table2} corresponds to universal NSI, that is, flavor independent interactions. When we do not take into account the tensor interaction, we are left with four coefficients: $C^{V,LL}_{sc\ell\nu},C^{V,RL}_{sc\ell\nu},C^{S,RR}_{sc\ell\nu}$ and $C^{S,LR}_{sc\ell\nu}$. Notice that equations (\ref{gamma_leptonic},\ref{gamma_semileptonic}) have a different dependence on the Wilson coefficients, hence, when combining the leptonic decay rates and the semileptonic decay rates it is possible to extract a bound for each parameter even if we analyze the four parameters at a time. We have computed the allowed values for those universal coefficients at 95\% C.L. by varying
the four parameters at-a-time, i.e. those are the most general cases, this is because both scalar and vector universal NSI
may affect the Brs. On the other hand, we have also estimated the allowed regions by varying only one parameter
at a time (right column Table \ref{table2}), this is when only one lepton flavor independent NSI contributes to the physical process.

\item{\it Flavor dependent NSI}
Some models may induce only vector, as well as only scalar NSI at a time. As we will show in the next section,
the left-right model or the two Higgs doublet model are examples of each type of NSI, respectively.
In those cases, we can obtain the bounds for the corresponding Wilson coefficients.
Those coefficients may depend on the flavor of the lepton involved. Since we have only six $\mathcal{B}^{th}_i$s,
we can perform the $\chi^2$ analysis only if we assume scalar NSI or vector NSI at a time. In each case, for
the electron NSI, we use the channels $i=1,3$ and the CLEO data points from the kinematic distribution, for the muon $i=2,4,6$ and for the tau, only a fit can be performed with $i=5$;
channel $i$ as shown in Table \ref{table1}.
Results for both cases, scalar and vector flavor dependent NSI are listed table \ref{table3}.
As we have mentioned, those constraints can be applied to the THDMs. In those cases, Wilson
coefficients are positive. Hence, we have constrained Wilson coefficients either assuming they are
real positive numbers or just real numbers. We will show the effectiveness of those constraints for
specific models.
\end{itemize}

\subsection{Complex Wilson coefficients}
\begin{figure}
\includegraphics[width=0.9\textwidth]{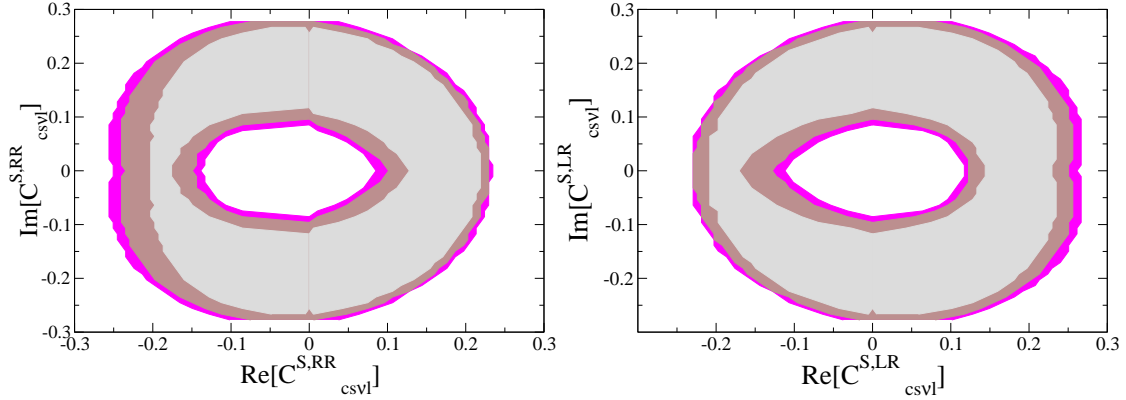}
\caption{Universal scalar NSI, parametrized by the complex coefficients 
$C^{S,rr}_{sc\ell\nu},C^{S,LR}_{sc\ell\nu}$, allowed from D meson decays. Colored
regions correspond to 68\% , 90\% and 95\% C.L respectively }\label{fig2b}
\end{figure}

\begin{figure}
\includegraphics[width=0.9\textwidth]{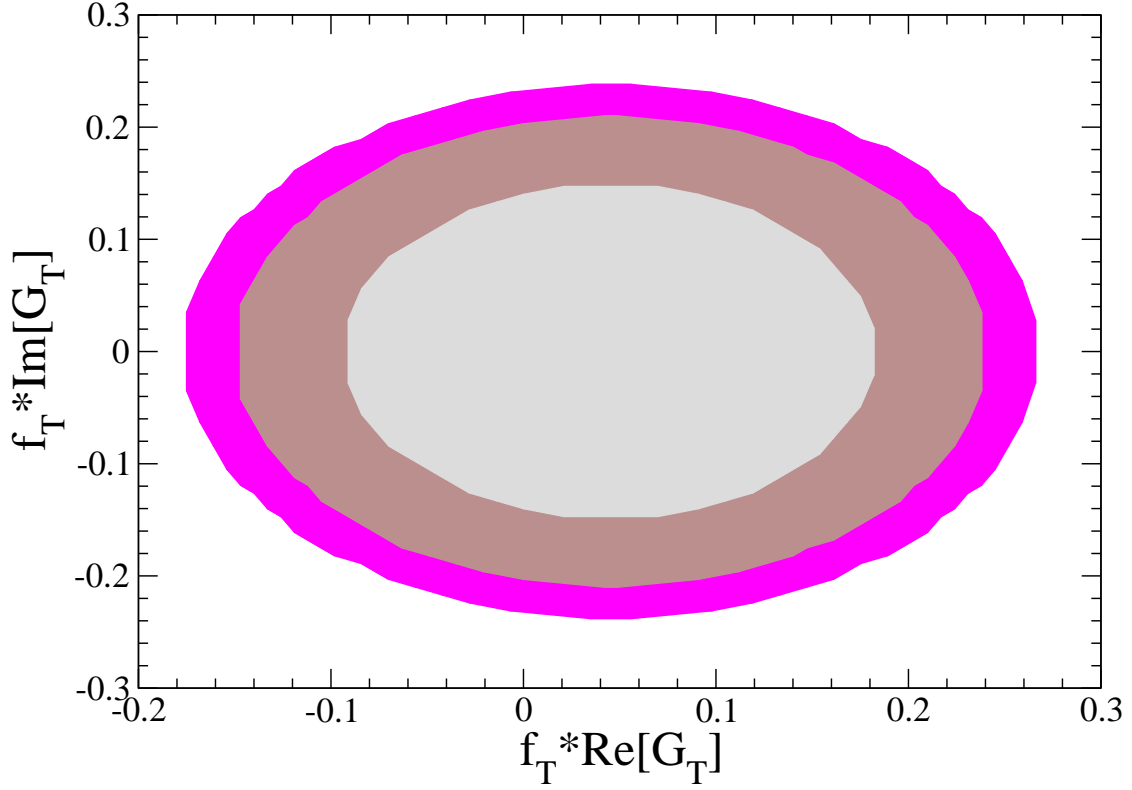}
\caption{Contrary to scalar or vector NSI, tensor NSI can not be
separated from the unknown form factor $f_T(0)$. Hence, we can only obtain the respective bounds for $Re[f_T G_T]$ and $Im[f_T G_T]$. Colored
regions correspond to 68\% , 90\% and 95\% C.L respectively}\label{fig2c}
\end{figure}

\begin{figure}
\includegraphics[width=0.9\textwidth]{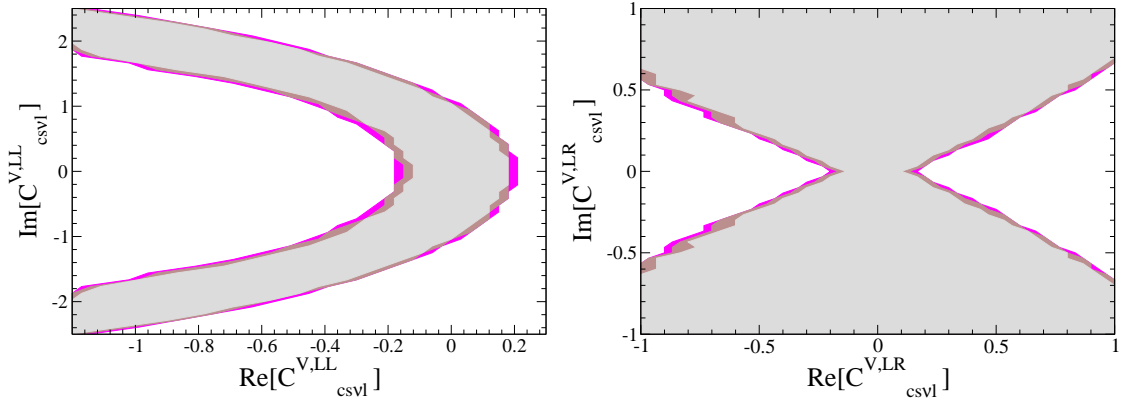}
\caption{Universal vector NSI, parametrized by the complex coefficients 
$C^{V,LL}_{sc\ell\nu},C^{V,RL}_{sc\ell\nu}$, allowed from D meson decays. Colored
regions correspond to 68\% , 90\% and 95\% C.L respectively}\label{fig2a}
\end{figure}

We shall consider now complex flavor universal Wilson coefficients. Many models of New Physics introduce CP violating phases which are in general not aligned with the SM CP violating phase, therefore we also analyze such scenario. Here, we assume that only one non-standard operator is dominant besides the Standard Model operator, either scalar, vector or tensor NSIs. This means we will take into account only one complex Wilson coefficient at a time, i.e. two independent parameters for each operator. We consider again a combined analysis of the leptonic and semileptonic BRs and the experimental data
from CLEO. \\
The model independent constraints at 68\%, 90\% and 95\% confidence level are shown in Figures \ref{fig2a},\ref{fig2b},\ref{fig2c}. 
Contrary to the scalar or vector NSI, tensor NSI can not be
separated from the unknown form factor $f_T(0)$. Hence, we can only obtain the bounds for $Re[f_T G_T]$ and $Im[f_T G_T]$, shown in Figure \ref{fig2c}. 
In summary, the allowed regions at 95\% C.L. are the following:
\begin{itemize}
\item vector NSI: $\chi^2/\mbox{d.o.f.}=0.96$
\begin{eqnarray}
-0.5<\mbox{Re}[C^{V,LL}_{sc\ell\nu}]<0.21, &\qquad& \mbox{95\% C.L.}\,, \nonumber \\
-1.63<\mbox{Im}[C^{V,LL}_{sc\ell\nu}]<1.63, &\qquad& \mbox{95\% C.L.}\,, \nonumber \\
-0.9<\mbox{Re}[C^{V,RL}_{sc\ell\nu}]<0.7, &\qquad& \mbox{95\% C.L.}\,, \nonumber \\
-2.1<\mbox{Im}[C^{V,RL}_{sc\ell\nu}]<2.1, &\qquad& \mbox{95\% C.L.}\,.
\end{eqnarray}
\item Scalar NSI: $\chi^2/\mbox{d.o.f.}=1.20$
\begin{eqnarray}
-0.24<\mbox{Re}[C^{S,RR}_{sc\ell\nu}]<0.23, &\qquad& \mbox{95\% C.L.}\,, \nonumber \\
-0.28<\mbox{Im}[C^{S,RR}_{sc\ell\nu}]<0.28, &\qquad& \mbox{95\% C.L.}\,, \nonumber \\
-0.23<\mbox{Re}[C^{S,LR}_{sc\ell\nu}]<0.26, &\qquad& \mbox{95\% C.L.}\,, \nonumber \\
-0.29<\mbox{Im}[C^{S,LR}_{sc\ell\nu}]<0.29, &\qquad& \mbox{95\% C.L.}\,.
\end{eqnarray}
\end{itemize}

We use the best fit points to compute the partial decays of the D meson, $D^{+} \to \bar{K}^{0} e^{+} \nu_{e}$ and $D^{0} \to K^{-} e^{+} \nu_{e}$, and we show them in Figure \ref{partialdecay}, compared with the experimental data and the Standard Model prediction. For those points we see there is better agreement with the experimental data.

\subsection{Transverse polarization estimation}
\begin{figure}
\includegraphics[width=.45\textwidth]{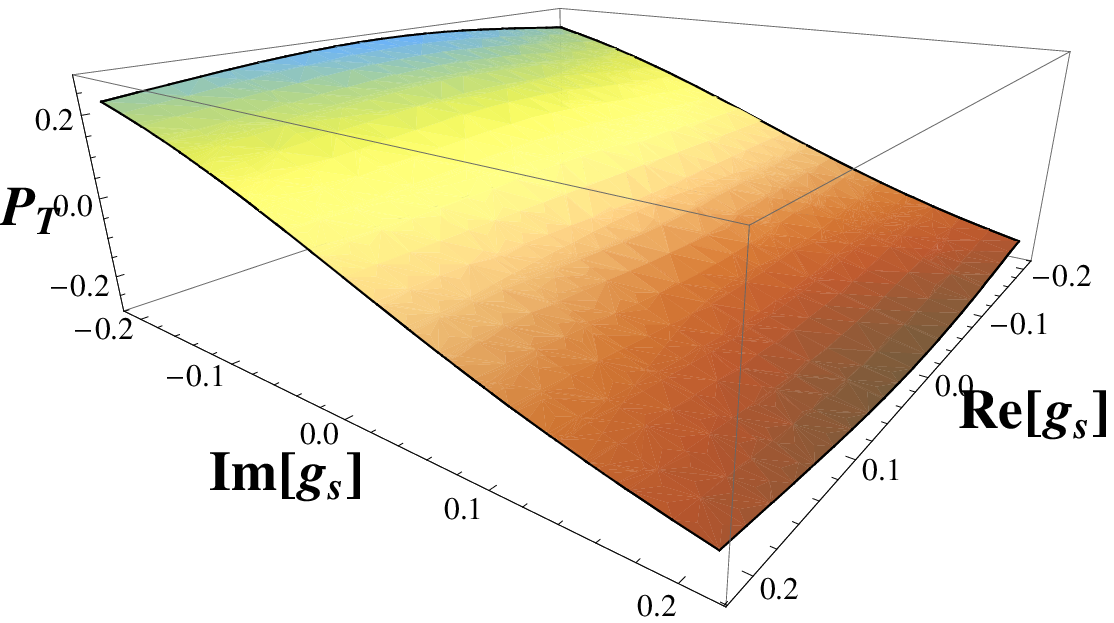}
\includegraphics[width=.45\textwidth]{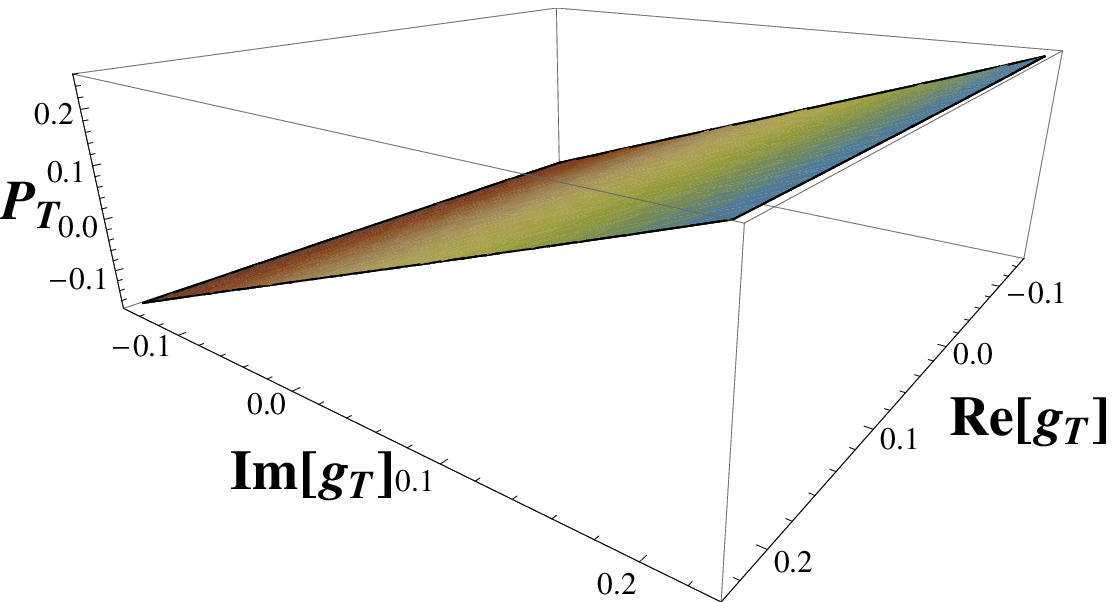}
\caption{Estimated transverse polarization $P_{T}$ for the $D^+\to \bar{K}^0 \ell^{+}\nu_\ell$. the left figure shows the $P_{T}$ when only scalar non-standard interactions are considered. The right figure is the $P_{T}$ when only tensor non-standard interactions are considered. The largest value of $P_{T}$ allowed from the previous constraints over the complex universal Wilson coefficients is $P_{T}=0.23$}\label{polarization}
\end{figure}

As an application of our results we give a prediction for a T-odd observable, the transverse polarization of the charged lepton for the decay $D^+\to \bar{K}^0 \ell^{+}\nu_\ell$. This observable has not been measured.  We chose this semileptonic decay thinking the experimental measurement could be done as in the case of the $K^{+}$ meson, \cite{Abe:2006de}. The $K^{+}$ decays as $K^{+}\to \pi^{0} \ell^{+} \nu_{\ell}$ and the BR of the $\pi^{0}\to \gamma\gamma$ is $BR(\pi^{0}\to \gamma\gamma)=98.823\pm0.034$\% \cite{
Beringer:1900zz}, this allows for a clean distinction of the angular distribution of the charged lepton, hence the transverse polarization. In our case, the $\bar{K}^{0}$ decays with a BR of $BR(\bar{K}^{0}\to\pi^{0}\pi^{0})=30.69\pm 0.05$\% \cite{
Beringer:1900zz}, allowing possibly for a distinction of the angular distribution of the charged lepton. In the SM it is identically zero which implies that a non zero value is a clear signal
of new physics. As we performed the analysis for the complex universal Wilson coefficients taking into account only one dominant non-standard operator the transverse polarization (\ref{pol}) can only be computed for each case. Furthermore, notice that if in the the transverse polarization (\ref{pol}) we only take into account the vector contribution it will vanish. For these reason we show the only non-vanishing transverse polarizations including New Physics integrated over all the kinematical allowed region. The results are shown in Figure (\ref{polarization}). \\
We can see in Figure (\ref{polarization}) that there is little dependence on the real contributions of both scalar and tensor non-standard interactions. The largest value of the transverse polarization allowed from the previous constraints over the complex universal Wilson coefficients is $P_{T}=0.23$, which is not negligible.

\section{Model Dependent Analysis:}\label{s4}
Let us consider now different models of New Physics. We perform a $\chi^{2}$ analysis in a model dependent way by finding the respective bounds over
the relevant parameters for those models. In particular, we obtain bounds for the Two Higgs Doublet Model Type-II and Type
III, the Left-Right model, the Minimal Supersymmetric Standard Model
with explicit R-Parity violation and Leptoquarks. We show that under some
simplifying assumptions, the model independent constraints can be mapped to
some particular models, exemplifying the usefulness of this kind of analysis.\\

\subsection{Two Higgs doublet model (THDM):}
\begin{center}
\begin{figure}
\includegraphics[width=0.8\textwidth]{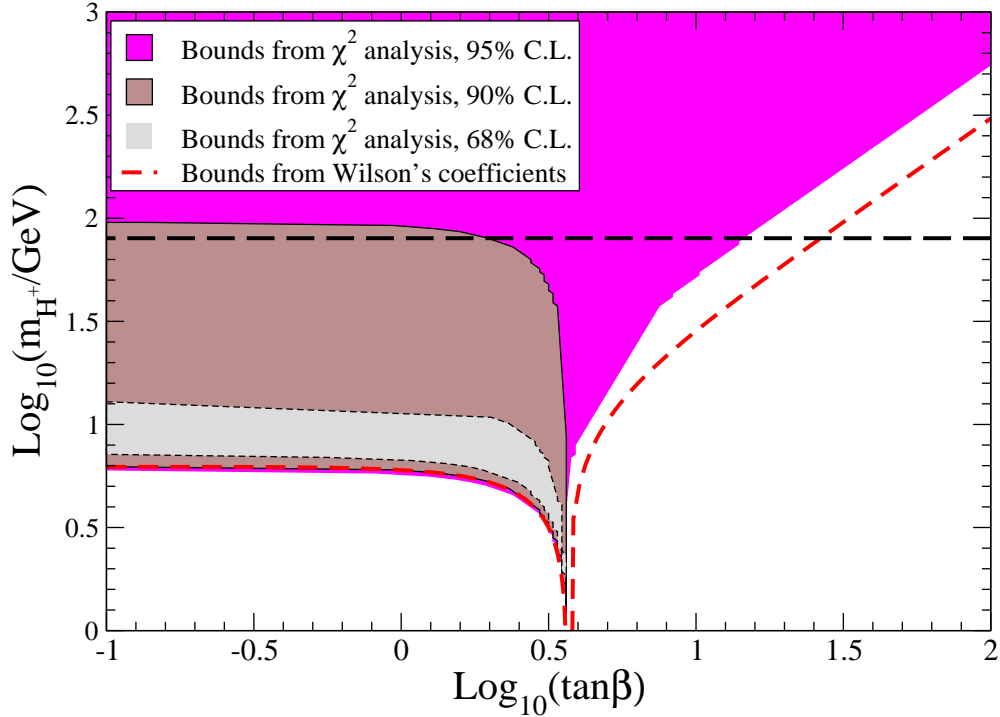}
\caption{Allowed regions for $\tan \beta$ and the mass of the charged Higgs to be consistent
with the D meson decays at 68\% C.L., 90\% C.L. and 95\% C.L. obtained performing a complete
$\chi^2$ analysis of the BRs. Dashed lines are the limits at 90\% C.L. using the bounds on
Wilson coefficients (Table \ref{table3}) showing good agreement. As a reference,
the LEP limit on the mass of a charged Higgs is also plotted \cite{Abbiendi:2013hk}.
}\label{fig3}
\end{figure}
\end{center}

It is one of the simplest and economical
extensions of the SM, see\cite{Lee:1973iz,Branco:2011iw} for a review.
THDM introduces an additional scalar doublet that induces scalar charged currents ($H^\pm$),
two neutral scalar fields and a pseudoscalar neutral field ($h^0,H^0$ and $A^0$).
For D meson decays, the  only two parameters involved are
the new scalar mass ($m_{H^+}$) and the ratio of the vacuum expectation values $\tan{\beta}$ of the two Higgs doublets.
At low energies, the Lagrangian for THDM, in the Higgs basis for the charge scalars and the mass basis for fermions,
is given by \cite{Barger:1989fj}
\begin{eqnarray}
-\mathcal{L}_{H^\pm}&=&\sqrt{2}/v H^+[V_{u_id_j}\bar{u}_i(m_{u_i}XP_L+m_{d_j}YP_R)d_j+ \nonumber \\
&+&m_\ell Z\bar{\nu}_L\ell_R]+\textrm{H.c.}
\end{eqnarray}
with $X,Y,Z$ functions of $m_{H^+}$ and $\tan{\beta}$ different for different versions of THDM, and the Wilson
coefficients will be given by
\begin{equation}
\frac{C^{S,RR}_{sc\ell\nu}}{2\sqrt{2}}=V_{cs}^*\frac{m_\ell m_c}{M_H^2}ZX\,,\quad
\frac{C^{S,LR}_{sc\ell\nu}}{2\sqrt{2}}=V_{cs}^*\frac{m_\ell m_s}{M_H^2}ZY\,.\label{WilsonTHDM}
\end{equation}
In particular, THDM-II
has Natural Flavour Conservation, namely the suppression of Flavor Changing Neutral Currents (FCNC) at tree level,
through a $Z_2$ symmetry \cite{Glashow:1976nt}. Interesting bounds have been obtained with meson decay experiments
\cite{Akeroyd:2009tn} and recently LEP has reported a lower bound on the mass of the charged Higgs of 80 GeV
\cite{Abbiendi:2013hk}.\\
For THDM-II, $X=\cot\beta$, $Y=Z=\tan\beta$. We perform a $\chi^{2}$ analysis using the 26 observables from the leptonic and semileptonic BRs and the kinematical distribution from CLEO. The result is shown in Figure \ref{fig3}. We can see from this figure that D meson decays favor lower masses for the charged Higgs at 90\% C.L., however at 95\%, there is good agreement with LEP bounds.\\
Now we will illustrate the effectiveness of our model independent bounds, once we apply them to Wilson coefficients of THDM, eqs (\ref{WilsonTHDM}).
There is a flavor dependence coming from the mass of the leptons involved.
Since this is an scalar interaction, we can use the bounds on flavor dependent scalar NSI.
From $C^{S,RR}_{sc \tau \nu_\tau}-C^{S,LR}_{sc\tau\nu_\tau}$ we get the region $-1.8\times 10^{-3}~\mbox{GeV}^{-1}<(m_c-m_s\tan^2\beta)/M_H^2<0.023~\mbox{GeV}^{-1}$
at 68\% C.L., which gives the outer region of an ellipse and the inner region of an hyperbole in the plane $(m_H,\tan\beta)$
illustrated in Fig. \ref{fig3}.
Those regions are in excellent agreement with the region
obtained by a complete $\chi^2$ analysis performed with all $D$ meson decays. The allowed values for $\tan \beta$ and $m_H^+$
are plotted in Fig. \ref{fig3} in a shadow gray area.  This
agreement illustrates the effectiveness of using generic Wilson coefficient to constrain the relevant
parameters of models beyond the SM.
A more complete analysis for the THDM-II model using different observables from flavor physics has been done in \cite{Deschamps:2009rh}.
Our aim in this work is not to compete with those constraints, rather than illustrate the effectiveness of this type of generic analysis
and to show that semileptonic decays may shed complementary information.
For completeness, let us briefly mention the THDM-III case which can be
analyzed immediately by noting that the Wilson coefficients in this case correspond to the following definitions:
\begin{eqnarray}
X & =& \cot\beta-\frac{\csc\beta}{\sqrt{2\sqrt{2}G_F}}m_c^{-1}\left(\tilde{Y}_{1,22}^u+\frac{V_{us}}{V_{cs}}\tilde{Y}_{1,21}^u+\frac{V_{ts}}{V_{cs}}\tilde{Y}_{1,23}^u\right), \\
Y&=& \tan\beta-\frac{\sec\beta}{\sqrt{2\sqrt{2}G_F}}m_s^{-1}\left(\tilde{Y}_{2,22}^d+\frac{V_{cd}}{V_{cs}}\tilde{Y}_{2,12}^d+\frac{V_{cb}}{V_{cs}}\tilde{Y}_{2,32}^d\right),\\ 
Z&=& \tan\beta-\frac{\sec\beta}{\sqrt{2\sqrt{2}G_F}}m_\ell^{-1}\tilde{Y}_{2,\nu_\ell\ell}^\ell \\
\end{eqnarray}
where $\tilde{Y}^f_{a,ij}$ are the Yukawa elements as were defined in \cite{DiazCruz:2004pj,DiazCruz:2009ek}. The corresponding bounds obtained via
eqs. \ref{WilsonTHDM} for THDM-III are interesting since they show relations between $\tilde{Y}^f_{a,ij}$, $\beta$ and
the mass of the charged Higgs.\\

\subsection{Left-right model:}
\begin{figure}
\includegraphics[width=0.9\textwidth]{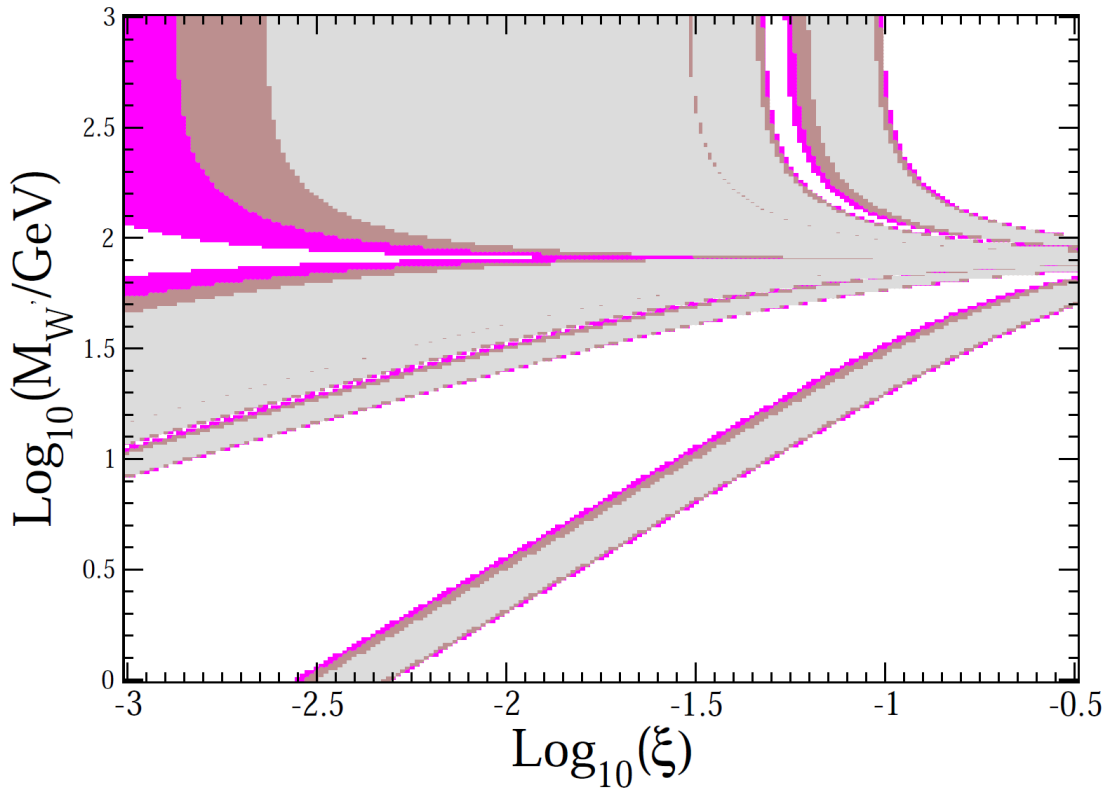}
\caption{Bounds at 68\% ,90\%  and 95\% C.L. on $\xi$ and the mass of the $W'$ boson obtained by using D
meson decays data.}\label{left_right}
\end{figure}

As an example of a model with vector NSI as the main contribution to new physics, we will consider SM's extensions based on extending the SM gauge group including a gauge $SU(2)_R$. The original model, based on the gauge group $SU(3)_C\times SU(2)_L\times SU(2)_R\times U(1)_{B-L}$,  restores the parity symmetry at high energies \cite{Pati:1973rp,Mohapatra:1974hk,Mohapatra:1974gc,Senjanovic:1975rk,Senjanovic:1978ev}. This SM extension has been extensively studied in previous works (see for instance refs.\cite{Beall:1981ze,Cocolicchio:1988ac,Langacker:1989xa,Cho:1993zb,Babu:1993hx}).  Recent bounds on the mass of $W_R$\cite{Beringer:1900zz,Alexander:1997bv,Acosta:2002nu,Abazov:2006aj,Abazov:2008vj,Abazov:2011xs,Chatrchyan:2012meb,Chatrchyan:2012sc,Aad:2011yg,Aad:2012ej} have strongly constrained these models. TWIST collaboration \cite{MacDonald:2008xf,TWIST:2011aa} found a model independent limit on $\xi$ to be smaller than 0.03 (taking $g_L=g_R$ at muon scale) through precision measurements of the muon decay parameters.
However the presence of right-handed currents may weaken some tensions between inclusive and exclusive determinations of some CKM elements \cite{Crivellin:2009sd,Chen:2008se,Feger:2010qc}, so it is appropriate to explore less restrictive versions of the LR model. Recently and ample phenomenological analysis has been done for the LR model using B physics \cite{Blanke:2011ry}, nevertheless we shall see that current D physics can shed complementary bounds on the free parameters of the model for a specific scenario. Here we consider the scenario where Left-Right is not manifest, that is $g_L\neq g_R$ at unification scale, with the presence of mixing between left and right bosons through a mixing angle $\xi$. This LR mixing is restricted by deviation to non-unitarity of the
CKM quark mixing matrix.  In case of manifest LR model, it is well known that $\xi$ has to be smaller than 0.005\cite{Wolfenstein:1984ay} and $M_{W'}$ bigger than 2.5 TeV\cite{Maiezza:2010ic}. But in the no manifest case, the constraint on $M_{W'}$ are much less restrictive as $M_{W'}$ could be as 
light as $0.3$ TeV\cite{Olness:1984xb} and $\xi$ can be as large as $0.02$\cite{Langacker:1989xa,Jang:2000rk,Badin:2007bv,Lee:2011kn}. 
The Lagrangian for this case, including only the vertex of interest, is given by
\begin{eqnarray}
-\mathcal{L}_{LR}&=&\frac{g_L}{\sqrt{2}}\bar{u}_i\gamma^\mu\left[\left(c_\xi V_{u_id_j} P_L+\frac{g_R}{g_L}s_\xi \bar{V}^R_{u_id_j}P_R\right)W_{\mu}^+\right.\nonumber\\
&&+\left.\left(-s_\xi V_{u_id_j}P_L+\frac{g_R}{g_L}c_\xi\bar{V}^R_{u_id_j}P_R\right){W'}_{\mu}^+\right]d_j\nonumber\\
&&+\frac{g_L}{\sqrt{2}}\bar{\nu}_L\gamma^\mu\left(c_\xi W^+_{\mu}-s_\xi {W'}_{\mu}^+\right)\ell_{L}+\textrm{H.c.},
\end{eqnarray}
where $c_\xi=\cos\xi$ and $s_\xi=\sin\xi$ and $W^+$, ${W'}^+$ are the mass states of gauge bosons. Likewise $\bar{V}^R_{u_id_j}=\exp^{-i\omega}V^R_{u_id_j}$, where $\omega$ is a CP violating phase.
After integrating degrees of freedom in a usual way, this leads to the Wilson coefficients
\begin{eqnarray}
C_{sc\ell\nu}^{V,LL}&=&\sin^2\xi \left(\frac{M^2_{W}}{M_{W'}^2}-1\right)V_{cs}\label{CVLL}\\
C_{sc\ell\nu}^{V,RL}&=&\frac{g_L}{2g_R}\sin(2\xi)\left(1-\frac{M_W^2}{M_{W'}}\right)\bar{V}^R_{cs}
\end{eqnarray}
Such scenarios were studied for instance in \cite{Delepine:2012xw}. In our case, the relevant parameters are: $\xi$, $M_{W'}, g_L/g_R \mbox{Re}[V_{cs}^R]$ and $g_L/g_R \mbox{Im}[V_{cs}^R]$.
By performing the combined analysis for all our 26 observables, by varying these four parameters at a time we
found the allowed regions for $\xi$ and  $M_{W'}$ which are shown in Fig. \ref{left_right}. 
There is only one viable restriction for the following parameters: $-71.0<g_L/g_R \mbox{Re}[V_{cs}^R]<83 $, while the analysis is insensitive to the imaginary part.

\subsection{MSSM-$\cancel{R}$:} 
R-Parity is a discrete symmetry defined as $(-1)^{3B+L+2S}$, where $B,L$ and $S$ are the baryon number, lepton number and particle spin respectively. R-Parity violating (RPV) interactions involve either lepton number violation or baryon number violation, but not both in order to preserve proton stability. These interactions lead to flavor violating interactions in the
leptonic and hadronic sector, and read explicitly as, 
 \begin{eqnarray}
 \mathcal{L_S}&=&\lambda_{ijk}\left[\tilde{\nu^{i}}_{L}\overline{e^{k}_{R}}e^{j}_{L}+\tilde{e^{j}}_{L}\overline{e^{k}_{R}}\nu^{i}_{L}+\tilde{e^{* k}}_{R}\overline{(\nu^{i}_{L})^{c}}e^{j}_{L}-(i\leftrightarrow j)\right]\nonumber \\
 &+& \lambda'_{ijk}\left[\tilde{\nu}^{i}_{L}\overline{d^{k}_{R}}d^{j}_{L}+\tilde{d}^{j}_{L}\overline{d^{k}_{R}}\nu^{i}_{L}+\tilde{d^{*}}^{k}_{R}\overline{(\nu^{i}_{L})^{c}}d^{j}_{L}- \tilde{e}^{i}_{L}\overline{d^{k}_{R}}u^{j}_{L}\right.\nonumber \\
&-&\left.\tilde{u}^{j}_{L}\overline{d^{k}_{R}}e^{i}_{L}-\tilde{d^{*}}^{k}_{R}\overline{(e^{i}_{L})^{c}}u^{i}_{L}\right]\,+\,h.c.
 \end{eqnarray}
A vast majority of observables have been used to set the corresponding bounds to these effective couplings (for a complete review see \cite{Barbier:2004ez} and references therein); in particular, for D meson decays \cite{Aida:2010xi,Cho:2011bd,Akeroyd:2002pi,Dreiner:2001kc,Dreiner:2006gu,Kundu:2008ui,Kao:2009fg,Bhattacharyya:2009hb,Akeroyd:2010qy,Tahir:2011ca,Davidson:1993qk}. We relate the corresponding Wilson coefficients constrained in the global analysis to the RPV couplings which constructively interfere with the Standard Model, i.e. through the exchange of a $-1/3$ electrically charged squark in a t-channel, which fixes the neutrino flavor, described by,
\begin{eqnarray}
\mathcal{L}_{S}&=&V^{*}_{cs}\frac{\sum_k|\lambda_{i2k}'|^{2}}{m^{2}_{\tilde{d^{k*}_R}}}(\bar{\nu_L}^{i}s^ {c}_R\bar{l^ {c}_R}c_L)\nonumber \\
&\underrightarrow{\text{ \tiny Fierz }}&V^{*}_{cs}\frac{\sum_k|\lambda_{i2k}'|^{2}}{2m^{2}_{\tilde{d^{k*}_R}}}(\bar{\nu_L}^{i}\gamma^{\mu}l^{i}_{L}\bar{s_L}\gamma_{\mu}c_L),
\end{eqnarray}
where  a Fierz transformation is done to rearrange the former operator in terms of the product of a  leptonic and a hadronic current.
The only non-vanishing Wilson coefficient is
\begin{equation}
C^{V,LL}_{sc\ell\nu}=\sqrt{2}V_{cs}/G_F\sum_k|\lambda_{i2k}'|^{2}/m^{2}_{\tilde{d^{k*}_R}} \label{vecsusy}
\end{equation}
Using the conservative bounds for the model independent constraints (table \ref{table3}) we get the following
constraints at $95\%$ confidence level and expressed in GeV$^{-2}$:
\begin{eqnarray}
0.05<\sum_k|\lambda_{12k}'|^{2}/(m^{2}_{\tilde{d^{k*}_R}}/300\mathrm{GeV})< 0.11, \nonumber \\
\sum_k|\lambda_{22k}'|^{2}/(m^{2}_{\tilde{d^{k*}_R}}/300\mathrm{GeV})< 0.17, \nonumber \\
\sum_k|\lambda_{32k}'|^{2}/(m^{2}_{\tilde{d^{k*}_R}}/300\mathrm{GeV})< 0.22
\label{boundssusy}
\end{eqnarray}
Our bounds agree with those found in \cite{Kao:2009fg} for muon and tau flavor. Nevertheless it is interesting to note that for the electron flavor we find more restrictive bounds. This is taking into account the latest and more accurate values of the form factors from lattice QCD as previously mentioned.

\subsection{The effective leptoquark lagrangian}
 Leptoquark particles are scalars or vectors bosons that carry both baryon number and lepton number \cite{Pati:1974yy},\cite{Buchmuller:1986zs}. These new particle states are expected to exist in various extensions of the SM. Leptoquark states usually emerge in grand unified theories (GUTs) \cite{Langacker:1980js,Frampton:1991ay,Hewett:1988xc} (as vector) or technicolor models (as scalars) \cite{Farhi:1980xs,Lane:1991qh}, or SUSY models with R-parity violation (as we saw in the previous section), 
but are described naturally in low energy theories as an effective for fermion interaction of a more fundamental theory. Effective interactions induced by leptoquark exchange can be manifest in meson decays, in particular, for the second generation of quarks in D meson decays. A vast majority of observables have been used to set the corresponding bounds to these effective couplings; in particular, for D meson decays \cite{Davidson:1993qk,Dobrescu:2008er,Dorsner:2009cu}.
 Leptoquarks are usually classified by their appropiate quantum numbers under the gauge symmetries of the Standard Model, such as colour, hypercharge, and isospin charge \cite{Buchmuller:1986zs}. These particles may couple to both quark chiralities, the left handed or right handed, in particular the scalar leptoquark $S$. When we rearrange the effective interactions in order to have external quark and lepton currents we do some Fierz transformations that lead to tensor, scalar and vector interactions, that we shall take into account in a model dependent analysis. We will consider the exchange of the scalar leptoquarks: $S_{0}$ with charge $-1/3$ and $(3,1,-2/3)$ gauge numbers; and the $S_{1/2}$ with charge $2/3$ and $(3,2,7/3)$ gauge numbers. Hence, the effective Lagrangian for the $c\to s$ transition (Fig.\ref{NSIgeneric}) is given by
 \begin{eqnarray}
 L^{LQ}_{Eff}&=&V^{*}_{cs}\left[\frac{\kappa^{R*}_{i2}\kappa^{L}_{i2}}{m^{2}_{S_{1/2}^{2/3}}}(\overline{\nu^{i}_L}c_R\overline{l^ {c}_{iL}}s^{c}_R)\right. \nonumber \\ &+& \left.  \frac{\kappa'^{R*}_{i2}\kappa'^{L}_{i2}}{m^{2}_{S_{0}^{-1/3}}}(\overline{\nu^{i}_L}s^{c}_R\overline{l^ {c}_{iL}}c_R)+\frac{|\kappa'^{L}_{i2}|^{2}}{m^{2}_{S_{0}^{-1/3}}}(\overline{\nu^{i}_L}s^{c}_R\overline{l^ {c}_{iR}}c_L)\right]
 \end{eqnarray}
After a Fierz transformation we have, in terms of the Wilson operators,
\begin{align}
\mathcal{L}^{LQ}_{Eff}&=\frac{1}{2}V^{*}_{cs}\left[\left(\frac{\kappa^{R*}_{i2}\kappa^{L}_{i2}}{m^{2}_{S_{1/2}^{2/3}}}-\frac{\kappa'^{R*}_{i2}\kappa'^{L}_{i2}}{m^{2}_{S_{0}^{-1/3}}}\right)\left(\overline{\nu_L}^ {i}l_{iR}\overline{s_L}c_R- \right.\right.\nonumber \\
-& \left.\left. \frac{1}{4}\overline{\nu_L}^ {i}\sigma_{\mu\nu}l_{iR}\overline{s_L}\sigma^{\mu\nu}c_R\right) 
- \frac{|\kappa'^{L}_{i2}|}{m^{2}_{S_{0}^{-1/3}}}\left(\overline{\nu}^{i}\gamma^{\mu}P_L l_{i} \overline{s}\gamma_{\mu}P_L c\right) \right]
\end{align}

Hence the only non-vanishing Wilson coefficients are $C^{VLL}_{sc\ell\nu},C^{VLR}_{sc\ell\nu},C^{SRR}_{sc\ell\nu}=-4C^{TRR}_{sc\ell\nu}$, given by,
\begin{equation}
C^{TRR}_{sc\ell\nu}=\frac{\sqrt{2}V_{cs}}{G_F}\left(\frac{\kappa^{R*}_{i2}\kappa^{L}_{i2}}{m^{2}_{S_{1/2}^{2/3}}}-\frac{\kappa'^{R*}_{i2}\kappa'^{L}_{i2}}{m^{2}_{S_0^{-1/3}}}\right)
\end{equation}

\begin{equation}
C^{VLL}_{sc\ell\nu}=-\frac{\sqrt{2}V_{cs}}{G_F}\left(\frac{|\kappa'^{L}_{i2}|}{m^{2}_{S_0^{-1/3}}}\right)\label{vector}
\end{equation}

Where the last Wilson coefficient also derives from the SUSY \cancel{R} Lagrangian. In the following we show the respective bounds as a result from our $\chi$ analysis considering the 26 observables: the leptonic and semileptonic decays of the D meson and the CLEO data points of the $q^{2}$ distribution. Notice here that we have one complex and one real independent Wilson coefficients (as the tensor operator is proportional to the scalar operator), hence the model dependent analysis is done varying 3 parameters at a time. At 95\% C.L. and expressed in GeV$^{-2}$ these are given by:

\begin{eqnarray}
-0.17<\mathrm{Re}\left(\kappa^{R*}_{i2}\kappa^{L}_{i2}-\kappa'^{R*}_{i2}\kappa'^{L}_{i2}\right)/(m^{2}_{S_0}/300\mathrm{GeV})<0.01 , \nonumber \\
-0.09<\mathrm{Im}\left(\kappa^{R*}_{i2}\kappa^{L}_{i2}-\kappa'^{R*}_{i2}\kappa'^{L}_{i2}\right)/(m^{2}_{S_0}/300\mathrm{GeV})< 0.10, \nonumber \\
0.04<|\kappa'^{L}_{i2}|/(m^{2}_{S_0}/300\mathrm{GeV})<0.11
\end{eqnarray}

As an example we can consider the leptoquark states that couple to the second generation of left handed quarks (chiral generation leptoquarks) and the first generation of left handed leptons. This means we take into account only the coefficient in Eq. (\ref{vector}), which also derives from the SUSY \cancel{R} effective lagrangian and corresponds to Eq. (\ref{vecsusy}). Therefore the Wilson coefficient is real and flavor dependent on the first generation of leptons, hence, we use the model independent constraints obtained in Section (\ref{s3}) for flavor dependent parameters, given in Table (\ref{table3}) which correspons to the first constraint in Eq. (\ref{boundssusy}). The allowed region at 95\% C.L. from the semileptonic decays of the D mesons is given by:

\begin{eqnarray}
0.05\left(\frac{m^{2}_{S_0}}{300\mathrm{GeV}}\right)<|\kappa_{12}'|^{2}< 0.11\left(\frac{m^{2}_{S_0}}{300\mathrm{GeV}}\right)
\end{eqnarray}

Previous bounds \cite{Beringer:1900zz} for the second generation of left handed quarks couplking to the first generation of left handed leptons, are reported to be $\kappa'^{2}<5.0\times(M_{LQ}/300\mathrm{GeV})$ for $S_0$. As stated in the previous subsection (for the MSSM-\cancel{R}), for the electron flavor and the second generation of quarks, this former constraint is more restrictive than previous bounds. 

\section{Conclusions}\label{s5}
We have constrained non standard interactions using D meson decay processes. We have combined the $D_s \to \ell \nu_\ell$ and the semileptonic  $D^0 \to K\ell\nu_\ell$ and $D^{+} \to K\ell\nu_\ell$ decays, together with the $q^ {2}$ distribution of the semileptonic decays for the electron channel measured by CLEO. The theoretical BRs were computed with the latest
lattice results on $f_+(q^2)$ and $f_0(q^2)$ form factors \cite{Koponen:2013tua}.
We have found the corresponding bounds for
the Wilson coefficients that parameterize the contribution of new physics as non standard interactions. We considered two scenarios in which the New Physics models have either aligned or not aligned physical phases with those of the SM, i.e. real or complex Wilson coefficients.
Those constraints can be applied to some model of new physics generating scalar, vector, or tensor operators, such as the THDM-II, Type III, the Left-Right model, MSSM-$\cancel{R}$ and leptoquarks, which we analyzed here. We show the usefulness of the model independent constraints as well as specific cases when a model dependent analysis is needed. In our model dependent analysis we found that for the THDM-II a low mass for the charged Higgs is favored, at 90\% C.L. $6.3\mathrm{GeV} <m_{H^{+}}<63.1\mathrm{GeV}$. We showed there are no strong restrictions for the LR model with these combined decay channels but comparable with previous bounds \cite{Olness:1984xb,Langacker:1989xa,Jang:2000rk,Badin:2007bv,Lee:2011kn}. In particular for the MSSM-\cancel{R}, our bounds agree with those found in \cite{Kao:2009fg} for muon and tau flavor. Nevertheless it is interesting to note that for the electron flavor we find more restrictive bounds. For the leptoquark model, taking into account the couplings to the second generation of left handed quarks and first generation of left handed leptons the constraints coming from D meson decays result encouraging if compared with previous bounds \cite{Beringer:1900zz}. This is taking into account the latest and more accurate values of the form factors from lattice QCD as previously mentioned.\\
We estimated the transverse polarization and found that these model independent constraints obtained from the D meson decays allow a large $P_T$, which is expected to be zero in the SM. Hence the experimental measure of $P_T$ could be useful to constraint Wilson coefficients involving quarks of the second generation.




\subsection*{Acknowledgements}
This work has been supported by CONACyT SNI-Mexico. The authors are also grateful to Conacyt (M\'exico) (CB-156618), DAIP project (Guanajuato University) and PIFI (SEP, M\'exico) for financial support.

\end{document}